\documentclass[lettersize,journal]{IEEEtran}
\usepackage{amsmath,amssymb,amsfonts,amsthm}
\usepackage{mathtools}
\usepackage{algorithm}
\usepackage{algpseudocode}
\usepackage{array}
\usepackage[caption=false,font=normalsize,labelfont=sf,textfont=sf]{subfig}
\usepackage{xcolor,makecell}
\usepackage{textcomp}
\usepackage{multirow}
\usepackage{stfloats}
\usepackage{xcolor}
\usepackage{todonotes}
\usepackage{url}
\usepackage{verbatim}
\usepackage{nameref}
\usepackage{graphicx}
\usepackage{dsfont}
\usepackage{cite}
\usepackage{subcaption}
\usepackage{hyperref}
\usepackage{wrapfig} 
\usepackage{booktabs}
\usepackage{orcidlink}
\newcommand{\cell}[2]{\begin{tabular}[c]{@{}c@{}}#1\\(#2)\end{tabular}}
\newcommand{\RMSE}{\operatorname{RMSE}}

\begin{document}

\pagestyle{empty}
\title{Resilient Decentralized Ergodic Coverage for Scalable Multi-Robot Systems in Unknown Time-Varying Environments}

\author{Maria G. Mendoza~\orcidlink{0000-0002-5285-5719},Victoria Marie Tuck~\orcidlink{0009-0009-4806-4744},Chinmay Maheshwari~\orcidlink{0000-0003-3596-2851}, and S. Shankar Sastry~\orcidlink{0009-0000-9021-7235}%
\thanks{M. G. Mendoza is with the Department of Mechanical Engineering,
        University of California, Berkeley, CA 94720, USA (e-mail: maria\_mendoza@berkeley.edu). V. M. Tuck is with the Department of Electrical and Systems Engineering,
        University of Pennsylvania, Philadelphia, PA 19104, USA (e-mail: vtuck@upenn.edu). C. Maheshwari is with the Department of Electrical and Computer Engineering,
        Johns Hopkins University, Baltimore, MD 21209, USA (e-mail: chinmay\_maheshwari@jhu.edu). S. Sastry is with the Department of Electrical Engineering and Computer Science,
         University of California, Berkeley, CA 94720, USA (e-mail: sastry@coe.berkeley.edu).}%


}


\maketitle
\thispagestyle{empty}

\begin{abstract}
Maintaining situational awareness in high-stakes multi-robot applications requires balancing exploration of unobserved regions with sustained monitoring of important regions, often in unknown, time-varying environments, under partial observability and degraded operating conditions. We propose a decentralized, graph-based ergodic coverage framework for persistent, long-horizon monitoring as a high-level planner. The environment is represented as an \textit{importance map}: a distribution over a graph of regions whose weights define how important each region is, with regions of interest (ROIs) denoting those of high importance. Each agent maintains an online Gaussian Process (GP) belief over this map from its own noisy observations and those of its neighbors, and follows an ergodic Markov-chain policy that steers its long-run visitation toward this belief and is updated locally as the belief changes. Unlike existing approaches that assume known importance maps or dynamics, centralized coordination, or a static environment, our framework addresses these challenges jointly. We evaluate the framework in long-horizon UAV disaster-response simulations in which ROIs spread, appear, and relocate, and analyze its computational complexity and belief accuracy. We show that the team keeps covering the map while consistently finding new and shifting ROIs, remains resilient to communication and memory degradation and robot loss, and scales to hundreds of robots.
\end{abstract}


\section{Introduction}

\IEEEPARstart{D}ecentralized multi-robot coverage in unknown, time-varying environments is used in many autonomous applications: environmental monitoring, persistent surveillance, search and rescue (SAR), and disaster response. These missions are challenging because the spatial distribution of importance, or \emph{importance map}, containing the regions of interest (ROIs), is unknown, evolves under unknown dynamics, and must be inferred online from sparse, noisy sensing.

As a motivating example, consider disaster response, where mission success depends on real-time situational awareness requiring continuous monitoring of evolving hazards and victim locations under harsh conditions. Unmanned Aerial Vehicles (UAVs) are well-suited to these settings~\cite{UAV-SAR-survey-lyu}, motivating ``Drone as a First Responder (DFR)" programs across U.S. emergency agencies \cite{DFR-AIAA}. Yet, most deployments remain human-supervised, limiting how many UAVs can be operated without diverting personnel from tactical response. This calls for decentralized autonomy, which lets the team scale with its size and remain resilient under \emph{degraded conditions}: intermittent or delayed communication, limited onboard memory, and the loss of robots mid-mission. Search in this setting is fundamentally a \emph{coverage} problem: sensing effort must be allocated across the environment in proportion to each region's importance.
Classical coverage path planning visits every region and terminates once coverage is complete~\cite{coverage-Choset2001}; disaster response instead requires \emph{persistent} coverage, revisiting regions throughout the mission. Standard ergodic coverage provides this persistence by matching the team's long-run visitation to a target distribution~\cite{ergodic-exploration-miller,real-time-md-ergodic}, but assumes this target is known. In our setting, the target distribution is unknown and evolves without a known model, so it must be learned online from the data the robots themselves collect.

Many existing multi-robot coverage and search frameworks rely on assumptions misaligned with these settings: centralized coordination or reliable communication, a static task structure, or known importance maps and environmental dynamics (Sec.~\ref{sec:related}). Each addresses part of the problem, but none at once sustains persistent coverage, learns an unknown, time-varying map without a model of its evolution, and scales to large teams without central coordination. The closest work~\cite{distributed-mr-ergodic-time-varying} proposed a distributed ergodic controller for time-varying environments, but each robot follows the gradient of a smoothed potential over a continuous domain, and its tracking guarantee assumes the map changes smoothly. In disaster response, critical regions may change abruptly, and neighboring regions can differ arbitrarily. This motivates the central question of our work:

\textbf{\emph{Can a large, decentralized team of robots allocate its coverage across regions according to their evolving importance in unknown, highly time-varying environments?}}

\begin{figure}
    \centering
    \includegraphics[width=1\linewidth]{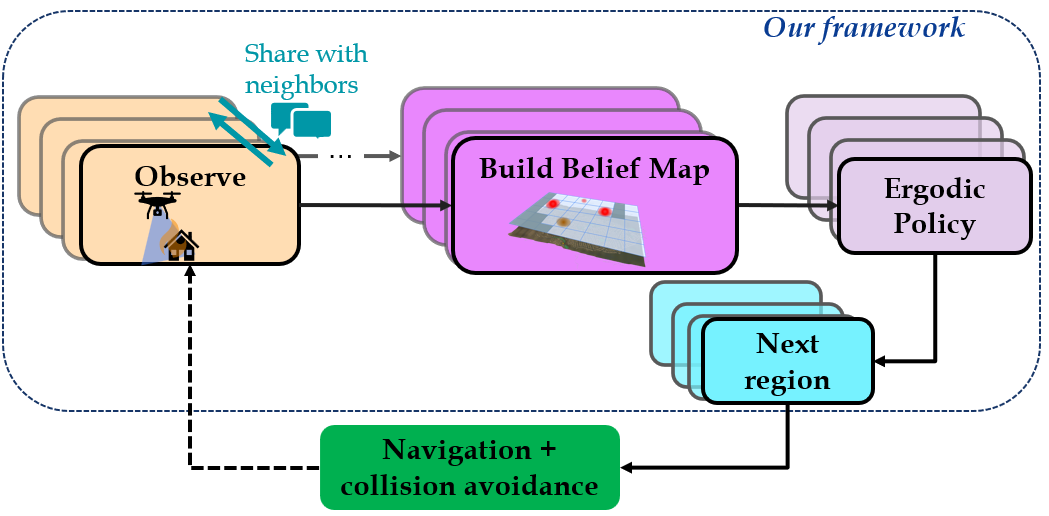}
    \caption{Framework overview. Each robot senses a region in the unknown, time-varying environment, fuses its own and its neighbors' observations into a belief map (Gaussian Process), and uses it to build an ergodic Markov-chain policy. The planner decides which region to visit next, and navigation and collision avoidance are handled by a low-level controller.}
    \label{fig:framework}
\end{figure}

We address this with a decentralized, graph-based ergodic planner in which each robot builds its policy online from local observations, so that the team's long-run visitation of each region matches a target distribution encoding how important every region is~\cite{ergodic-theory-calvinMoore} (Fig.~\ref{fig:framework}). Since this distribution is unknown, each robot maintains a Gaussian Process (GP) to build a belief over the unknown map, updated online from its own noisy observations and those of its neighbors, and computes a policy that drives its visitation toward a belief target distribution.

\textbf{\textit{Our main contributions}} are: (i) a decentralized, graph-based ergodic coverage framework for persistent, long-horizon coverage of unknown, time-varying distributions, operating as a regional planner in which each robot updates its own policy locally as its belief of the environment changes; ii) complexity analysis and ablation studies showing that computation, memory, and communication costs do not grow with team size or horizon; (iii) extensive simulations in UAV disaster-response scenarios across different team sizes and time-varying environments, including a 200-robot study under degraded communication, limited memory, and robot attrition; (iv) comparisons to ergodic and non-ergodic baselines showing lower regret, faster on average ROI discovery, and more frequent full map coverage.

\section{Related Works}\label{sec:related}

We position prior work along three requirements of long-term continuous monitoring. (i) A \emph{persistent} objective that guides robots to spend time in regions proportional to their importance, rather than a one-shot coverage that ends once the workspace is covered or converges to a static configuration. One-shot coverage falls short when (ii) the importance field is \emph{unknown and time-varying} because it must be inferred online without a model of its evolution (e.g., prescribed dynamics or a parametric form), whose changes are detected only by revisiting. And lastly, (iii) \emph{decentralized} operation that scales to large teams, with no central planner, and can maintain the mission with local communication or robot loss.

\textbf{\textit{Persistent vs.\ one-shot coverage.}} Most coverage methods are one-shot: they terminate once the workspace is covered or the field is mapped~\cite{zhou2023racer, LCD-RIC-2024}, converge to a static configuration~\cite{yazicioglu2017, distributed-coverage-control-time-varying}, or avoid resampling visited regions~\cite{scalar-field-adaptive-nav-Fuentes2025}. In autonomous search and monitoring, however, a target can appear in a region already explored. Methods that converge to a configuration can stall in local minima of the coverage cost, so they use mechanisms to promote exploration, such as increasing GP uncertainty~\cite{distributed-coverage-control-time-varying, GP-UCB-Dorfler-26} or re-planning at the end of each episode~\cite{Zhang2024mac-dt}. Sec.~\ref{sec:comparison} shows this is in time-critical missions; the policy has to also sustain exploration. Learning-based policies~\cite{coverage-control-gnn, DRL-MA-SAR-2023, coordinated-uavs-pd, sagr-language-guided-search, co-navgpt-vlm, diffvas-aamas26} incur substantial offline training and degrade when deployment departs from training~\cite{survey-rl-time-varying}. Other methods instead make persistence the objective by revisiting regions: Markov-chain surveillance~\cite{george2018markov, diaz2023distributed} and ergodic control~\cite{ergodic-theory-calvinMoore, ergodic-exploration-miller, real-time-md-ergodic, decentralized-ergodic-sensing, dec-erogdic-lygeros, naveed_adaptive_2025} shape long-run visitation so that time-averaged spatial statistics converge to a target distribution. In both, however, the target is fixed or specified through a known model.

\textbf{\textit{Unknown, time-varying fields.}} When the importance field is unknown, it must be inferred online, typically with GPs. Existing GP-based methods, however, treat the field as static. Some fuse neighbors' measurements into local GPs under limited communication~\cite{LCD-RIC-2024}; others improve the reconstruction of spatial fields, such as terrain and bathymetry, through kernels whose variability changes across space~\cite{chen2022ak} and constant-time sparse updates~\cite{chen2024poam}; and GP-based coverage learns a static density~\cite{Zhang2024mac-dt, GP-UCB-Dorfler-26}. Their non-stationarity, where addressed, is spatial rather than temporal. Methods where the field varies in time assume its evolution is known: known map dynamics~\cite{mo-ergodic-dynamic}, known time-varying densities~\cite{eberhard2025time, decentralized-ergodic-sensing}, known target motion~\cite{ma-ergodic-lowsensor-choset, Nikolaos-Gpappas-graphNN}, smoke fields simulated offline~\cite{ma-ergodic-timevaryingsensor-choset}, or known per-cell information-decay rates~\cite{naveed_adaptive_2025}. The few that infer a time-varying field without such a model either forget old data~\cite{bogunovic_time-varying_2016, distributed-coverage-control-time-varying} or treat the field as static over short periods and smooth enough to follow its estimated gradient~\cite{scalar-field-adaptive-nav-Fuentes2025}. We assume no model of the field's evolution, and therefore evaluate on maps with abrupt temporal changes (Sec.~\ref{sec:exp}).

\textbf{\textit{Decentralized coverage.}} Most ergodic methods solve a continuous trajectory optimization that couples high-level decisions (where to go) with low-level control (how to get there)~\cite{mo-ergodic-dynamic, ergodic-exploration-miller, ma-ergodic-timevaryingsensor-choset, dec-erogdic-lygeros}. Its decision variables are the states and controls at every point of every robot's trajectory, so when the team is optimized jointly the problem grows with both the horizon and the number of robots~\cite{ma-ergodic-timevaryingsensor-choset, dec-erogdic-lygeros}; it is non-convex, so solvers return local minima that depend on initialization that is either solved offline for a static distribution~\cite{dec-erogdic-lygeros} or solved at every replanning step~\cite{mo-ergodic-dynamic, ma-ergodic-timevaryingsensor-choset}. With multiple robots, the team's time-averaged statistics enter each robot's planner, so each robot maintains other robots' full trajectories~\cite{dec-erogdic-lygeros} or exchanges its Fourier coefficients with all others~\cite{real-time-md-ergodic} or with its neighbors through consensus over a connected network~\cite{decentralized-ergodic-sensing}. This motivates separating the two layers and discretizing the workspace into regions.


Graph-based formulations do this: an ergodic planner over a graph of regions selects the next region, and a lower layer generates the trajectory between them~\cite{shirose2024gesce, wong2025rapidly, wong2025annealing}. The planner then scales with the graph rather than the horizon or the team, and stays modular to different vehicles. Hardware experiments confirm that this decomposition transfers from simulation~\cite {wong2025rapidly, wong2025annealing}. These methods, however, plan for a single robot~\cite{shirose2024gesce, wong2025rapidly} or centrally synthesize the multi-robot chain~\cite {wong2025annealing}, and all assume a static field. We leverage this decomposition but construct a \textit{decentralized} approach that can adapt to a \textit{time-varying} environment.


Closest to ours, Mantovani et al.~\cite{distributed-mr-ergodic-time-varying} embed a GP-estimated density in heat equation-driven area coverage (HEDAC), where each robot follows the gradient of a heat-equation solution and the guarantee assumes a smoothly varying field. We instead plan over regions whose importance can differ arbitrarily between neighbors and change abruptly, so the MH policy uses only ratios of neighboring values, not gradients. To our knowledge, no prior method is at once persistent, learns an unknown time-varying field without a model of its evolution, and is decentralized at scale; we address all three with a \textbf{\textit{decentralized regional planner}}.

\section{Model and Problem Formulation}\label{sec:model-problem}
\subsection{Model}
\label{sec:model}

We consider a team of unmanned aerial vehicles (UAVs) tasked with tracking a time-varying spatial information distribution over a graph-based environment. The environment is modeled as an undirected graph \(\mathcal{G}=(\mathcal{R},\mathcal{E})\), where \(\mathcal{R}\) is a finite set of regions and \(\mathcal{E}\subseteq\mathcal{R}\times\mathcal{R}\) encodes adjacency. 
The neighborhood of region \(r\) is \(\mathcal{N}(r):=\{r'\in\mathcal{R}:(r,r')\in\mathcal{E}\}\).
Motion constraints, obstacles, and no-fly zones are encoded by removing infeasible nodes or edges from \(\mathcal{G}\).

Each region is sized to the sensing footprint of a UAV, and one time step \(k=0,1,\ldots\) is the time to transit between adjacent regions and sense the arrival region. Navigation between regions is delegated to the onboard low-level controller (Sec.~\ref{sec:exp-setup}).
Let \(\mathcal{M}:=\{1,\ldots,M\}\) denote the set of UAVs. At time \(k\), UAV \(m\in\mathcal{M}\) occupies region \(x_k^m\in\mathcal{R}\) and moves between regions according to \(x_{k+1}^m\in\mathcal{N}(x_k^m)\).  UAVs communicate within radius
\(R_{\mathrm{comm}}\); the communication model is specified in
Sec.~\ref{sec:observation-model}.

The mission-level sensing priority at time \(k\) is represented by an information map \(\phi_k^\star:\mathcal{R}\to\mathbb{R}_{\geq 0}\), where
\(\phi_k^\star(r)\) is the relative importance of region \(r\). In
disaster-response settings, high values may correspond to suspected human presence, smoke intensity, thermal signatures, active fire fronts, or other indicators requiring sustained monitoring. The \textbf{\textit{information map is time-varying}}, it evolves as \(\phi_k^\star=\mathcal{U}_k(\phi_{k-1}^\star)\), where \(\mathcal{U}_k\) is an unknown exogenous transformation which changes at times \(\mathcal{K}\). The UAVs know neither
\(\phi_0^\star\), \(\mathcal{K}\), nor \(\{\mathcal{U}_k\}\). Note that we do not assume any parametric dynamic evolution model.\footnote{The dynamics of the environment are left unstructured: they need not be smooth, and neighboring regions may carry arbitrarily different levels of importance. Sec.~\ref{sec:exp} describes the environment evolutions used in our experiments.}

Assuming \(\sum_{r\in\mathcal{R}}\phi_k^\star(r)>0\), the corresponding target spatial distribution is
\begin{equation}
\label{eq:desired-target-distro}
    \rho_k^\star(r)
    :=
    \frac{\phi_k^\star(r)}
    {\sum_{r'\in\mathcal{R}}\phi_k^\star(r')},
    \qquad r\in\mathcal{R}.
\end{equation}
That is, \(\rho_k^\star(r)\) is the fraction of long-run team visits that region \(r\) should receive if the map remained \(\phi_k^\star\). Because \(\rho_k^\star\) is defined region-wise, it does not restrict the number of sources: multiple fires or hazards produce multiple high-value regions in \(\phi_k^\star\).

The UAVs do not observe \(\phi_k^\star\) or \(\rho_k^\star\) directly. Instead, each UAV receives noisy measurements of the region it occupies,
\(r\in\operatorname{Ball}_{\mathcal{R}}(x_k^m,R_{\mathrm{sense}})\), UAV
\(m\) observes $y_k^m(r)=\phi_k^\star(r)+\varepsilon_k^m(r)$,
where the sensing noises are independent across UAVs, regions, and time, with \(\varepsilon_k^m(r)\sim\mathcal{N}(0,\sigma_\varepsilon^2)\). The local observation set of UAV \(m\) at time \(k\) is  
$Y_k^m := \bigl(x_k^m,\,k,\,y_k^m\bigr)$.

Each UAV maintains a belief \(\bar{\rho}_k^m\in\Delta(\mathcal{R})\) of the current target distribution, where \(\Delta(\mathcal{R})\) denotes the probability simplex over \(\mathcal{R}\). The belief is updated online from local and shared observations. We denote the team-averaged belief by
\begin{equation}
\label{eq:team-belief}
\bar{\rho}_k := M^{-1}\sum_{m=1}^M \bar{\rho}_k^m
\end{equation}
Because the target distribution varies over time, we measure physical coverage over a sliding window rather than the full horizon. Let \(W\) be the window length, \(\tau_k:=\max\{0,k-W+1\}\), and \(T_k:=k-\tau_k+1\). The empirical visitation distribution of UAV \(m\) is \(\hat{\rho}_k^m := T_k^{-1}\sum_{\tau=\tau_k}^{k} \mathds{1}_{\mathcal{R}}(x_\tau^m)\),  and the team's is its average,
\begin{equation}\label{eq:team-empirical}
\hat{\rho}_k := M^{-1}\sum_{m=1}^{M}\hat{\rho}_k^m.    
\end{equation} Thus, \(\hat{\rho}_k(r)\) is the fraction of recent team visits to region \(r\).

\subsection{Problem Formulation}
\label{sec:problem-formulation}
The objective is to design decentralized regional planning policies that use local sensing and local communication to track the unknown, time-varying target distribution \(\rho_k^\star\). Since \(\rho_k^\star\) is not directly observed, the UAVs must simultaneously estimate the information map from noisy local measurements and plan their region visits so that the empirical team visitation distribution \(\hat{\rho}_k\) matches \(\rho_k^\star\). The objective is \textbf{\textit{persistence}}; there is no terminal condition, and, rather than covering \(\mathcal{R}\) once, the team must keep reallocating its visits as \(\phi_k^\star\) evolves. Navigation between regions is left to each UAV's low-level controller and is outside the scope of this problem; in Sec.~\ref{sec:exp-setup}, we verify with a kinematic UAV model that the planned region sequences are executable at the chosen time step. 

We evaluate performance using two complementary criteria: physical tracking and belief accuracy. Physical tracking is measured by the instantaneous ergodic tracking error, the \(\ell_1\) distance between the team's empirical visitation \(\hat{\rho}_k\) in \eqref{eq:team-empirical} and the true target distribution \(\rho_k^\star\) in \eqref{eq:desired-target-distro}, 
$E_k:=\left\|\hat{\rho}_k-\rho_k^\star\right\|_1$
and by its time average over the horizon \(K\), the tracking regret,
\begin{equation}
\label{eq:regret}
    \operatorname{Regret}(K)
    :=
    \frac{1}{K}
    \sum_{k=1}^{K}
    \mathbb{E}
    \left[
        \left\|
            \hat{\rho}_k-\rho_k^\star
        \right\|_1
    \right],
\end{equation}
where the expectation is taken over sensing noise and algorithmic randomness, while the realized information process \(\{\phi_k^\star\}_{k=0}^K\) is treated as fixed, since it evolves exogenously and is unknown to the agents.

\section{Approach}
Our framework combines local information acquisition, GP-based belief updates, and ergodic policy synthesis. Algorithm~\ref{alg:ergodic-gp} summarizes the decentralized procedure, which each UAV operates separately. Each UAV starts from a uniform belief target distribution and an initial Metropolis-Hastings (MH) transition matrix. At each step, the UAV estimates the unknown information map from its own and neighbors' observations as a Gaussian Process and forms a GP-UCB belief target distribution (Sec.~\ref{sec:observation-model});  every \(\tau_P\) steps it updates its transition policy using Metropolis-Hastings whose visitation statistics track this target (Sec.~\ref{sec:MH}).

\subsection{Observation and Local Data Aggregation}
\label{sec:observation-model}
At time \(k\), UAV \(m\) records the observation of the region it occupies. UAVs exchange information only with neighbors within the communication radius,
\[
    \mathcal{N}_k^m
    :=
    \left\{
        \ell\in\mathcal{M}\setminus\{m\}:
        \|p(x_k^m)-p(x_k^\ell)\|_2\leq R_{\mathrm{comm}}
    \right\}
\]
which is symmetric since \(R_{\mathrm{comm}}\) is homogeneous across the team. At each step, UAV \(m\) shares \(Y_k^m\) with \(\mathcal{N}_k^m\) and updates its local dataset as
\begin{equation}
\label{eq:local-dataset-update}
    \mathcal{D}_k^m
    =
    \mathsf{Mem}\Bigl(
        \mathcal{D}_{k-1}^m
        \cup
        Y_k^m
        \cup
        \textstyle\bigcup_{\ell\in\mathcal{N}_k^m}Y_k^\ell
    \Bigr),
    \qquad
    \mathcal{D}_{-1}^m=\emptyset,
\end{equation}
where  $\mathsf{Mem}(\cdot)$ denotes the memory model. In the full-memory case, $\mathsf{Mem}$ is the identity map, so UAVs retain all previously collected or received observations. In memory-limited settings, $\mathsf{Mem}$ may retain only a finite window or bounded subset of past observations. Thus, communication remains local and instantaneous, while the memory model determines how much historical data each UAV uses for belief estimation.

\subsection{Gaussian Process Posterior and GP-UCB Belief Update}
\label{sec:build-unknown-map-GP}
Since the importance map \(\phi_k^\star\) and its evolution \(\{\mathcal{U}_k\}\) are unknown, each UAV estimates \(\phi_k^\star\) from \(\mathcal{D}_k^m=\{(z_i,y_i)\}_{i=1}^{n}\), \(n=|\mathcal{D}_k^m|\), with Gaussian process (GP) regression~\cite{rasmussen2006gaussian} over inputs \(z_i=(r_i,k_i)\) and observations \(y_i\) of region \(r_i\) taken at time \(k_i\), and converts the estimate into a target distribution with an upper confidence bound (UCB).\footnote{The GP is not a generative model of the environment; it provides an estimate and an uncertainty measure from the noisy observations in \(\mathcal{D}_k^m\).} 

We use a separable spatio-temporal kernel
\begin{equation}
\label{eq:kernel}
    \kappa(z,z') = \sigma_f^2\,
    C_{\nu_s}\!\bigl(d/\ell_s\bigr)\,
    C_{\nu_t}\!\bigl(|k-k'|/\ell_t\bigr),
\end{equation}
where \(d=\|p(r)-p(r')\|_2\) is the distance between the centroids \(p(r)\) of regions \(r\) and \(r'\), \(C_\nu\) is the Mat\'ern correlation with smoothness \(\nu\)~\cite{rasmussen2006gaussian}, and \(\ell_s,\ell_t>0\) are the spatial and temporal length-scales. The kernel is a prior for the estimator, not an assumption on \(\phi_k^\star\); Sec.~\ref{sec:ablation} compares spatial smoothness, temporal length-scale, and a joint (non-separable) alternative. The posterior mean \(\mu_k^m(r)\) and standard deviation \(\sigma_k^m(r)\) at \(z=(r,k)\) follow the standard GP equations~\cite{rasmussen2006gaussian} with noise variance \(\sigma_\varepsilon^2\). UAV \(m\) then forms the GP-UCB estimate and the belief target distribution


\begin{equation}
\label{eq:ucb-information-map}
    \bar{\phi}_{k,r}^m
    =
    \left[\mu_k^m(r)+\beta\sigma_k^m(r)\right]_+,
    \qquad r\in\mathcal{R},
\end{equation}
\begin{equation}
\label{eq:belief-target-distribution}
    \bar{\rho}_k^m(r)
    =
    \frac{\bar{\phi}_{k,r}^m}
    {\sum_{r'\in\mathcal{R}}\bar{\phi}_{k,r'}^m},
    \qquad r\in\mathcal{R},
\end{equation}
where \([a]_+=\max\{a,0\}\) and \(\beta>0\). The projection \([\cdot]_+\) keeps \(\bar{\rho}_k^m\in\Delta(\mathcal{R})\) even though individual measurements may be negative. The term \(\beta\sigma_k^m\) drives exploration and sustains persistence: as the time since a region was last observed grows, its posterior variance grows through \(\ell_t\), so regions not recently observed regain target mass and are revisited. The Markov-chain policy (Sec.~\ref{sec:MH}) uses only \(\bar{\rho}_k^m\).

\subsection{Ergodic Markov-Chain Policy Update}
\label{sec:MH}
Given the belief target distribution \(\bar{\rho}_k^m\), UAV \(m\) constructs a Markov-chain policy \(P_k^m\) over the graph using the Metropolis--Hastings (MH) rule. In particular, for a given target distribution $\bar{\rho}\in\Delta(\mathcal{R})$, we define a Markov kernel such that  for $r'\in\mathcal{N}(r)\backslash\{r\}$,  
\(
     P(r'\mid r)
    =
    \frac{1}{|\mathcal{N}(r)|} \min\left\{
        1,
        \frac{\bar{\rho}(r')|\mathcal{N}(r)|}
        {\bar{\rho}(r)|\mathcal{N}(r')|}
    \right\}
\)
and $P(r\mid r)
    =
    1-\sum_{r'\in\mathcal{N}(r)\backslash \{r\}}P(r'\mid r).$
This construction makes $\bar{\rho}$ the stationary distribution of $P$ (detailed balance). Since $\mathcal{G}$ is connected and $\bar{\rho}(r)>0$ for all $r\in\mathcal{R}$, $P$ is irreducible, and aperiodic. Hence, for a fixed $\bar{\rho}$, the UAVs' empirical visitation distribution converges to $\bar{\rho}$
\cite{levin2017markov}. To account for the time-varying importance map, $P_k^m$ is recomputed every $\tau_{P}$ steps, together with the belief
target $\bar{\rho}_k^m$.

\begin{algorithm}[t]
\caption{Decentralized Ergodic Exploration}
\label{alg:ergodic-gp}
\begin{algorithmic}[1]
\State \textbf{Input:} $T_{\mathrm{final}}$, $\tau_{\mathrm{P}}$, $R_{\mathrm{comm}}$, $R_{\mathrm{sense}}$, initial states $\{x_0^m\}_{m\in\mathcal{M}}$
\State \textbf{Initialize:} $\bar{\rho}_0^m \gets \frac{1}{|\mathcal{R}|}\mathbf{1}$, $\mathcal{D}_{-1}^m \gets \emptyset$, $P_0^m \gets \mathrm{MH}(\bar{\rho}_0^m)$ for all $m\in\mathcal{M}$

\For{$k=0,\ldots,T_{\mathrm{final}}-1$}
    \For{each UAV $m\in\mathcal{M}$}
        \State Collect $Y_k^m$ 
        \State Update $\mathcal{D}_k^m$  
        \If{$k \equiv 0 \pmod{\tau_{\mathrm{P}}}$}
            \State $\bar{\phi}_k^m,\bar{\rho}_k^m \gets \mathcal{GP}_{\mathrm{UCB}}(\mathcal{D}_k^m)$ (using \eqref{eq:ucb-information-map}-\eqref{eq:belief-target-distribution})
            \State $P_k^m \gets \mathrm{MH}(\bar{\rho}_k^m)$ 
        \Else
            \State $\bar{\rho}_k^m \gets \bar{\rho}_{k-1}^m$
            \State $P_k^m \gets P_{k-1}^m$
        \EndIf
        
        \State Sample $x_{k+1}^m \sim P_k^m(\cdot\mid x_k^m)$
    \EndFor
\EndFor
\end{algorithmic}
\end{algorithm}

\subsection{Complexity Analysis}
\label{sec:comp-complexity}
Table~\ref{tab:comp-cost} summarizes the per-agent computation complexity of the GP belief and policy, the memory complexity, and the communication complexity. The belief and policy are updated every \(\tau_P\) steps, while information sharing happens at every step. Long missions with many agents grow each agent's dataset $n = |\mathcal{D}_k^m|$, and an exact GP refit costs $O(n^3)$. Thus, we compare two ways to reduce the \(O(n^3)\) cost of the \emph{exact} GP, where \(n=|\mathcal{D}_k^m|\) and show an analysis in~Sec.\ref{sec:exp-belief}. Sparse GP regression (SGPR)~\cite{chen2024poam} approximates the \emph{kernel} with \(n_z\le n\) inducing points. Fusing compresses the \emph{data}: observations of the same region within a time bin of length \(\Delta_b\) are replaced by their inverse-variance mean \(\bar y=\bar\sigma^{2}\sum_i y_i/\sigma_i^{2}\), with \(\bar\sigma^{-2}=\sum_i\sigma_i^{-2}\), leaving \(n_f\le n\) points. The policy update scales with the number of edges, since MH computes transitions only between neighboring regions. Agents share only their new observations (row, column, time, value) with their \(\nu\) neighbors, not GP models or trajectories. With bounded memory, each agent keeps at most \(B\) recent observations, so \(n,n_z,n_f\le B\) and no term grows with team size or mission length. The memory needs scale only with $|\mathcal E|$ to store the transitions and $B$ for the bounded amount of data kept.

\begin{table}[t]
\caption{Complexity Analysis}
\label{tab:comp-cost}
\centering
\small
\setlength{\tabcolsep}{4pt}
\begin{tabular}{lccccc}
\toprule
 & \multicolumn{2}{c}{GP belief} & \multicolumn{2}{c}{MH policy} & Comm. \\
\cmidrule(lr){2-3}\cmidrule(lr){4-5}\cmidrule(lr){6-6}
 & Fit & Predict & Build \(P\) & Store & Share \\
\midrule
Exact & \(O(n^3)\)      & \(O(|\mathcal{R}|\,n^2)\)   & \multirow{3}{*}{\(O(|\mathcal{E}|)\)} & \multirow{3}{*}{\(O(|\mathcal{E}|+B)\)} & \multirow{3}{*}{\(O(\nu)\)} \\
SGPR  & \(O(n\,n_z^2)\) & \(O(|\mathcal{R}|\,n_z^2)\) & & \\
Fused & \(O(n_f^3)\)    & \(O(|\mathcal{R}|\,n_f^2)\) & & \\
\bottomrule
\end{tabular}
\end{table}



\section{Experiments} \label{sec:exp}
\begin{figure}[t]
    \centering
    \includegraphics[width=1.0\linewidth]{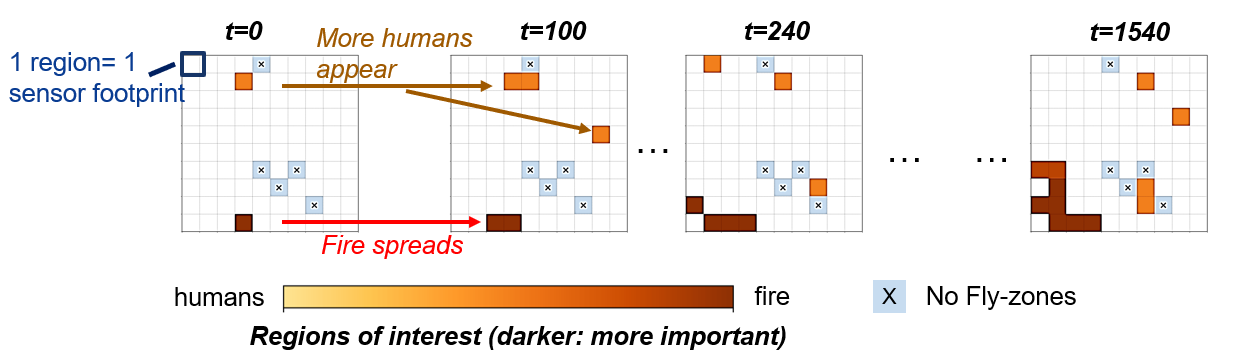}
    \caption{Example of a time-varying environment on a $10\times10$ region grid, where each region corresponds to one sensor footprint. The user specifies the relative importance of each type of region of interest (ROI): here, regions with evacuating humans have mid-level importance, and fire regions have high importance. Humans evacuate at different rates, so evacuee regions appear and disappear over time, while the fire ignites and spreads to neighboring regions.}
    \label{fig:scenario}
\end{figure}

\subsection{Setup and Metrics}\label{sec:exp-setup}

We evaluate the proposed decentralized ergodic coverage framework in a simulated disaster-response environment, discretized into a finite grid of regions with ROIs (Fig.~\ref{fig:scenario}). All simulations were implemented in Python\footnote{Code and configuration files to reproduce all experiments are available at \url{git@github.com:sastry-group/MAES.git}.} and run on a consumer laptop (Intel Core i7, 32\,GB RAM). We vary the map configuration (grid size, number and location of ROIs and no-fly zones) and its time-varying dynamics (rate of change), the team size (3–200 agents, including robot loss), and the communication model ($R_{\textrm{comm}}$ from one grid cell to global, with delays in information exchange).


\textbf{Space and Time Abstraction} Our framework targets large-scale multi-robot planning, so region size and $\Delta t$ are abstracted to the robot's sensing: a region is the area one robot can sense on a pass at its operating altitude, and a time step $\Delta t$ is the time needed to fly between adjacent regions and complete a detection pass. The same planner thus applies from small platforms such as Crazyflies to large first-responder UAVs; it assigns regions, not low-level trajectories. As an example platform for Drone-as-First-Responder operations, we use a Skydio X10 (Table~\ref{tab:kin-params}): at $120$\,m altitude, its camera covers about a $200$\,m region, and it reaches an adjacent region and completes a detection pass within $\Delta t=30$\,s.

\textbf{Time-varying environment model.} We model two types of change $\mathcal{U}_k$. First, spreading hazards (fire, smoke, contamination) \textit{spread} into neighboring regions, with spread rates taken from wildland-urban interface fire simulations~\cite{masoudvaziri2021swuift} ($0.06$--$0.49$\,m/s, i.e., $\approx 111$ to $14$ steps per region). Second, population-level demands (evacuees, congestion, requests for assistance) \textit{appear and disappear} as on/off regions with random dwell times. We do not assume a smooth field or track individual motion: a region is important when a hazard or demand is present.

\textbf{Flight dynamics.} To illustrate the high-level planner connected to a low-level controller, we show an animation\footnote{\url{github.com/sastry-group/MAES/tree/main/video}} of agents following double-integrator dynamics with position $\mathbf{p}_m\in\mathbb{R}^3$ and saturated velocity tracking~\cite{mueller2025dynamics},
\begin{equation}
\ddot{\mathbf{p}}_m = \mathrm{sat}_{a_{\max}}\!\Big(\tfrac{\mathbf{v}^{\mathrm{des}}_m - \dot{\mathbf{p}}_m}{\tau_v}\Big),
\end{equation}
with horizontal and vertical speed limits $v_{\max}$, $v_{z,\max}$. The desired velocity $\mathbf{v}^{\mathrm{des}}_m$ points to the assigned region's center, or follows a loiter orbit when the agent stays (an MH self-loop). Nearby agents are deconflicted by a repulsive velocity term and a small vertical split, with a $15$\,m separation target.

\begin{table}[t]
\centering
\caption{Physical parameters (Skydio X10 \cite{skydio_x10})}
\label{tab:kin-params}
\small
\begin{tabular}{lc}
\toprule
Region size / altitude & $200$\,m / $120$\,m \\
Time step $\Delta t$ (dwell) & $30$\,s ($3$\,s) \\
Max.\ horizontal / vertical speed & $16$ / $4$\,m/s \\
Max.\ acceleration & $3$\,m/s$^2$ (assumed) \\
Velocity time constant $\tau_v$ & $0.5$\,s \\
Loiter radius / speed & $60$\,m / $8$\,m/s \\
Separation target & $15$\,m \\
\bottomrule
\end{tabular}
\end{table}

\textbf{Metrics.} In addition to tracking regret~\eqref{eq:regret}, we also evaluate the belief error and the GP itself against the true map. We quantify the belief error as the \(\ell_1\) distance between the team-averaged belief \(\bar{\rho}_k\) \eqref{eq:team-belief}  and the true target distribution \(\rho_k^\star\) \eqref{eq:desired-target-distro},
\begin{equation}
\label{eq:belief-loss}
    \mathcal{L}_{\mathrm{belief},k}
    :=
    \left\|
        \bar{\rho}_k-\rho_k^\star
    \right\|_1.
\end{equation}
Thus, regret measures if the UAVs physically allocate sensing effort correctly, whereas \(\mathcal{L}_{\mathrm{belief},k}\) measures whether their beliefs match the current target distribution.

The root mean square error (RMSE) of the posterior mean
\begin{equation}
  \RMSE^m_k := \Big(\tfrac{1}{|R|}\textstyle\sum_{r\in R}
  \big(\mu^m_k(r)-\phi^\star_k(r)\big)^2\Big)^{1/2},
  \label{eq:rmse}
\end{equation}
isolates what the GP has learned about the field, independent of the exploration term $\beta\sigma^m_k$ and the normalization in \eqref{eq:team-belief}. The mean standardized log loss (MSLL)~\cite[Eq.~(2.34)]{rasmussen2006gaussian}
{\small\begin{equation}
  \mathrm{MSLL}^m_k := \tfrac{1}{|R|}\textstyle\sum_{r\in R}
  \Big[\tfrac12\log\!\big(2\pi\sigma^m_k(r)^2\big)
  + \tfrac{(\phi^\star_k(r)-\mu^m_k(r))^2}{2\sigma^m_k(r)^2}\Big]
  - \mathcal{L}_0,
\end{equation}}
evaluates $\sigma^m_k$. $\mathcal{L}_0$ is the loss of a trivial Gaussian predictor with the mean and variance of $\phi^\star_0$.
$\mathrm{MSLL}\le 0$ indicates calibrated uncertainty, and $\mathrm{MSLL}>0$ indicates overconfidence in the estimation.

We also report two event-based metrics. \emph{ROI discovery time} is the number of steps from when an ROI appears until an agent first visits it; ROIs that move or disappear before being visited count as missed. \emph{Cover time} is the number of steps for the team to visit every region at least once; over long horizons we report the number of complete covers and the average cover time (total steps divided by the number of covers).

\subsection{Belief Model: Kernel Ablation}\label{sec:ablation}

\begin{figure}
    \centering
    \includegraphics[width=1\linewidth]{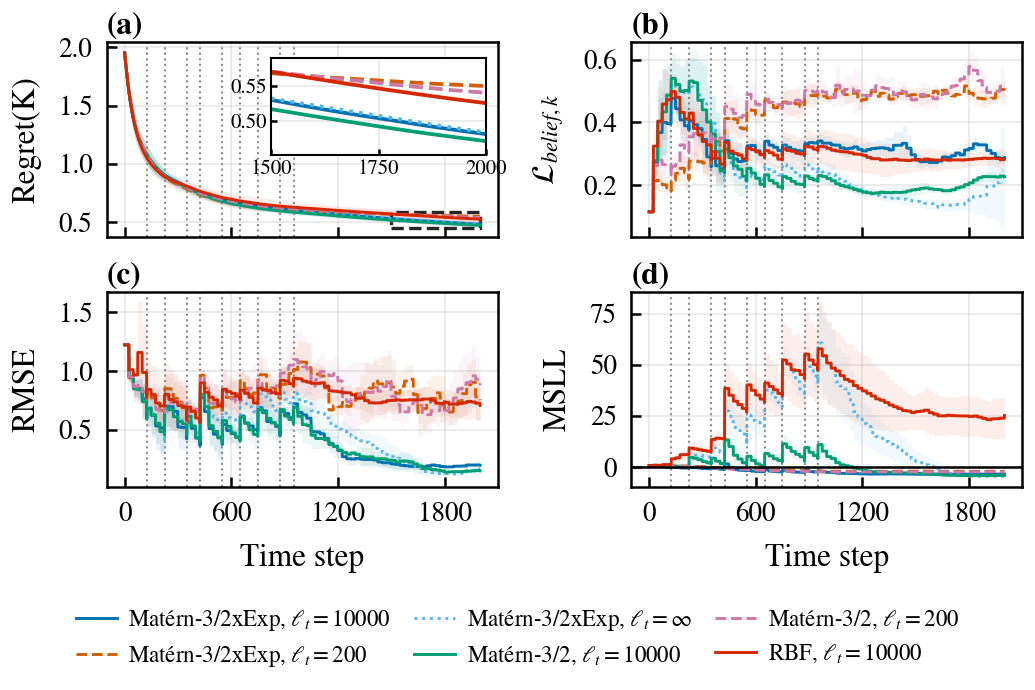}
    \caption{Kernel and temporal length-scale ablation.}
    (a) Regret (inset: final 200 steps), (b) belief error $\mathcal{L}_{belief,k}$,
    (c) RMSE, and (d) MSLL. Curves compare Matérn-3/2 and RBF kernels \cite{rasmussen2006gaussian} with the separable Matérn-3/2 $\times$ exponential kernel \cite{bogunovic_time-varying_2016} at temporal length scales $\ell_t \in \{200, 10^4, \infty\}$ 
    \label{fig:kernel_ablation}
\end{figure}

We ablate the GP kernel and its temporal length scale $\ell_t$, comparing the infinitely smooth radial basis function (RBF) kernel, a joint Mat\'ern-3/2 kernel over space and time, and a separable kernel that is Mat\'ern-3/2 in space and exponential in time~\cite{bogunovic_time-varying_2016}. The environment changes 10 times at irregular intervals and remains fixed after $t=1063$. The regret varies little across variants (Fig.~\ref{fig:kernel_ablation}a), so the differences lie in the belief (Fig.~\ref{fig:kernel_ablation}b--d). First, the spatial kernel must not be too smooth: at the region scale, the importance distributions are piecewise constant with sharp boundaries between ROIs and background, which the Matérn variants capture, while the RBF has among the highest RMSE, belief, and MSLL errors (Fig.~\ref{fig:kernel_ablation}c--d). Second, within the Matérn variants, $\ell_t$ governs how the belief tracks the changes, which we observe in RMSE and MSLL. With $\ell_t = 200$, the belief fails to learn the field. With $\ell_t = \infty$, pre-change observations are treated as current, so the variance stays small where the field has changed and the GP remains overconfident long after the environment settles. We also show the effect of the temporal kernel: with an exponential temporal factor, the variance grows with the time since the last visit. With $\ell_t = 10^4$, the separable kernel learns the field and stays calibrated during changes (Fig.~\ref{fig:kernel_ablation}c--d), at the cost of a slightly higher belief error than the kernels that retain old observations (Fig.~\ref{fig:kernel_ablation}b).

\subsection{Policy and Belief Map Updates}\label{sec:exp-tau}
\begin{figure}
    \centering
    \includegraphics[width=1\linewidth]{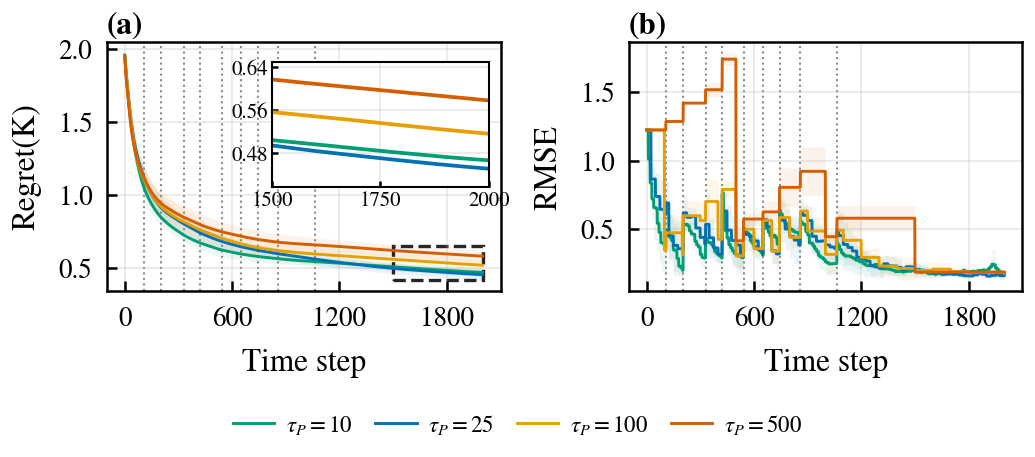}
    \caption{Effect of policy $P$ update frequency $\tau_P$. (a) Regret \eqref{eq:regret}; (b) RMSE \eqref{eq:rmse}}
    \label{fig:tau-update}
\end{figure}
We evaluate the effect of $\tau_{P}$, the update period for both the policy (MH) and the belief map, on performance. We use the same map configurations as in Sec.~\ref{sec:ablation} and the separable kernel with $\ell_t = 10^4$. Figure~\ref{fig:tau-update} reports the regret~\eqref{eq:regret} in (a) and the RMSE of the GP mean in (b). Performance is best with the smallest period and worst with the largest, because after an environment change, the target distribution is updated only at the next scheduled update, so the RMSE error remains high until then. However, as long as the update is fast enough relative to the environment dynamics, the exact value has little impact ($\tau_P = 10$ and $25$ perform similarly); from an application perspective, we show that the system does not require precise tuning of $\tau_P$.

\subsection{GP Belief Data Reduction}\label{sec:exp-belief}
\begin{figure}
    \centering
    \includegraphics[width=1.0\linewidth]{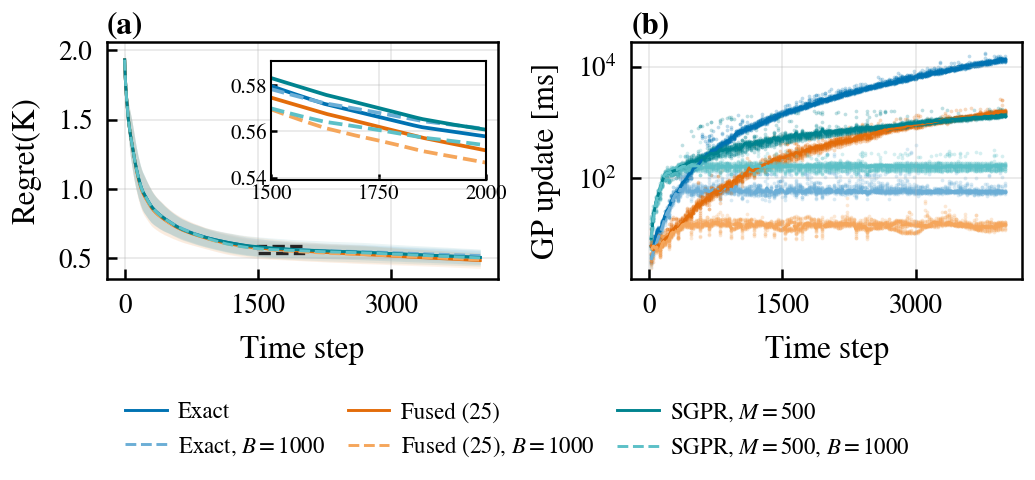}
    \caption{GP Belief models with full and bounded memory ($B=1000$): (a) regret (inset:1500-2000, during changes in the environment), (b) wall-clock time per GP update (fit and prediction).}
   
    \label{fig:gp-performance} 
\end{figure}

We compare SGPR and fusing (Sec.~\ref{sec:comp-complexity}) against the exact GP, each with full memory and with bounded memory (\(B=1000\)). Fusing uses \(\Delta_b=\tau_P=25\), which is nearly exact because it is much shorter than the interval between environment changes. To evaluate the growth with \(n\), we extend the horizon to \(T=4000\) steps on 3 maps with \(\sim\)10--15 changes each.
In Fig.~\ref{fig:gp-performance}a we show that data reduction does not affect the planner, the regret is nearly identical across methods. Fig.~\ref{fig:gp-performance}b shows the wall-clock time per update (fit plus prediction). With full memory, the cost of fusing grows with \(T\) at a rate similar to SGPR. With \(B=1000\), all three methods flatten, and fusing remains the cheapest, consistent with the bounded-memory costs in Table~\ref{tab:comp-cost}.


\subsection{Robustness to Communication Degradation and 
Attrition}\label{sec:exp-robust}
Table~\ref{tab:robust} reports ROI discovery and cover time for $N=6$ agents, without communication and with global communication, and up to four agents gradually lost within the first 900 steps. While losing agents will naturally slow ROI discovery and full map coverage, the system's performance does not significantly degrade. The slowdown grows sublinearly with the number of agents lost. Losing four of six agents ($2/3$ of the team) increases cover time by $1.4\times$ without communication and $1.6\times$ with global communication, which is below $3\times$ expected if cover time scaled inversely with team size. Similarly, ROI discovery roughly doubles. Results also show that communication between the two extremes does not degrade; our approach does not rely on communication to recover from attrition.

\begin{table}[t]
\centering\footnotesize
\setlength{\tabcolsep}{2pt}\renewcommand{\arraystretch}{0.95}
\caption{Effect of communication degradation and robot attrition: ROI discovery and map coverage time.}
\label{tab:robust}
\begin{tabular*}{\columnwidth}{@{\extracolsep{\fill}}lcccccc@{}}
\toprule
 & \multicolumn{3}{c}{ROI hit time} & \multicolumn{3}{c}{Cover time} \\
\cmidrule(lr){2-4}\cmidrule(lr){5-7}
Agents lost & 0 & 2 & 4 & 0 & 2 & 4 \\
\midrule
None & 31$\pm$10 & 42$\pm$16 & 67$\pm$18 & 228$\pm$58 & 283$\pm$58 & 329$\pm$51 \\
Global & 35$\pm$13 & 52$\pm$6 & 69$\pm$18 & 201$\pm$16 & 277$\pm$18 & 331$\pm$105 \\
\bottomrule
\end{tabular*}
\end{table}

\subsection{Scalability}\label{sec:exp-scale}

\begin{figure}
    \centering
    \includegraphics[width=0.9\linewidth]{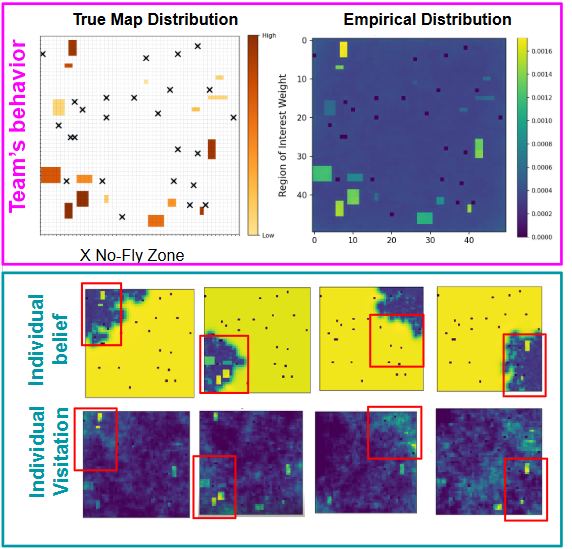}


    \caption{Large-scale simulation (200 agents, $50\times50$ grid, 40{,}000 steps). \textbf{Top:} team-level visitation versus the true target distribution. \textbf{Bottom: individual agent beliefs and {empirical visitation of four agents}, each covering a small portion of the environment (red box).}}
    \label{fig:scale-combined}
\end{figure}

We run a large-scale simulation with 200 agents on a $50\times50$ grid (2,500 regions) over 40,000 steps under severe communication and memory constraints. Each agent retains at most 4{,}000 past observations for its GP belief update, communicates only within $R_{\text{comm}} = 5.0$ (roughly 3\% of the grid), and shares with its neighbors observations from the past 100 steps only. Thus, no agent observes more than a small fraction of the graph, nor retains a complete history of observations. Even so, Fig.~\ref{fig:scale-combined} (top) shows that the team's time-averaged visitation captures the true importance map across the full environment. Fig.~\ref{fig:scale-combined} (bottom) each agent's belief is accurate only over a small, subregion, and its visitation targets that area. The division of labor emerges from local belief and local policy construction, without explicit partitioning, task assignment, or any agent reconstructing the global map. Since per-agent memory and communication do not grow with team size or horizon, the approach scales to settings where global planning is limited by computational cost.

\subsection{Comparison to baselines}\label{sec:comparison}
\begin{figure}
    \centering
    \includegraphics[width=1.0\linewidth]{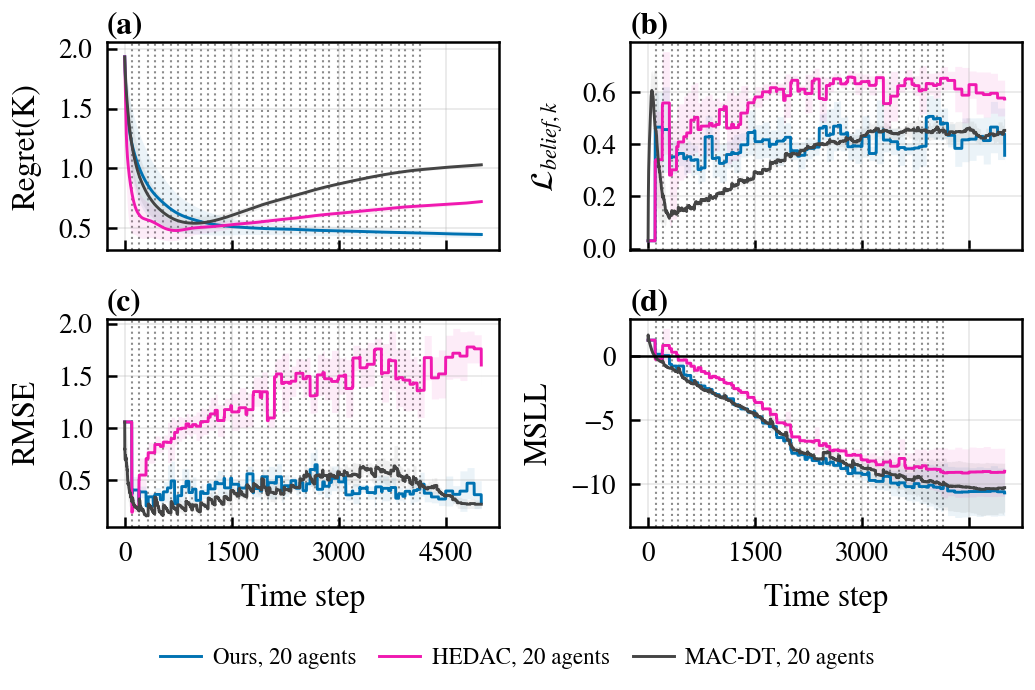}
\caption{Comparison to baselines: 20 agents on a $20\times20$ map with global communication. The environment changes 40 times, roughly every 100 steps (dotted lines), then remains fixed. (a) Regret, (b) belief error $\mathcal{L}_{\mathrm{belief},k}$, {\color{blue}{(c) RMSE, (d) MSLL.}}}
    \label{fig:mac-dt-ergodic-longterm}
\end{figure}

We compare against two baselines: MAC-DT~\cite{Zhang2024mac-dt}, a centralized GP-based coverage method that greedily steers agents toward high-reward regions, and HEDAC~\cite{distributed-mr-ergodic-time-varying}, a deterministic ergodic planner in which each robot follows the gradient of a heat-equation potential. Because MAC-DT is centralized, we compare all methods using global communication. While both baselines act faster initially, MAC-DT discovers ROIs first within the first 1k steps, and HEDAC's regret drops fastest (Fig.~\ref{fig:mac-dt-ergodic-longterm}a), they fail to be \textit{persistent} in the long horizon and to adapt to continuous environment changes. MAC-DT concentrates on regions already believed important, so its regret climbs and it explores the entire map with half the frequency (Table~\ref{tab:roi_discovery}). HEDAC is cheap per step, but its gradient-following settles into local minima of the potential, so agents stop revisiting regions and its RMSE and belief error grow (Fig.~\ref{fig:mac-dt-ergodic-longterm}b--c), it finds only 57\% of ROIs, and it explores the map less (4.5 covers vs 15.7 for ours). Our stochastic MH policy keeps revisiting the map, so its regret keeps decreasing after each change, and it finds 99\% of ROIs with less computational cost.

\begin{table}[t]
\centering
\caption{ROI discovery, coverage, and computation time.}
\label{tab:roi_discovery}
\footnotesize
\setlength{\tabcolsep}{4pt}
\begin{tabular}{lcccccc}
\toprule
 & \multicolumn{4}{c}{ROI discovery (success\%)} & Full cover & Compute \\
\cmidrule(lr){2-5}
Method & First 1k & All & High & Mid & (covers/run) & P/agent (ms) \\
\midrule
Ours   & 50.6 & \cell{\textbf{40.7}}{99\%}  & \cell{\textbf{42.3}}{100\%}  & \cell{\textbf{40.4}}{99\%}  & \cell{\textbf{351}}{$15.7$}  & \textbf{0.02} \\
MAC-DT & \textbf{38.2} & \cell{89.7}{85\%}  & \cell{127.0}{100\%} & \cell{80.0}{82\%}  & \cell{733}{$7.5$}   & 1.98 \\
HEDAC  & 226.7 & \cell{224.8}{57\%} & \cell{294.9}{98\%} & \cell{194.4}{48\%} & \cell{1222}{$4.5$} & 0.08 \\
\bottomrule
\end{tabular}
\vspace{2pt}
\begin{minipage}{\columnwidth}
\footnotesize
$N=20$ agents, two $20\times20$ maps, averaged over runs with different initial agent locations. Metrics as in Sec.~\ref{sec:exp-setup}; \emph{First 1k}: ROI discovery within the first 1000 steps; covers per run shown below cover time. \emph{Compute}: per-agent wall-clock time per planner step. Best per column in bold.
\end{minipage}
\end{table}


\section{Conclusion}
 This work presents a decentralized multi-robot coverage framework for unknown, time-varying environments under limited sensing and communication. By integrating ergodic control with online belief updates, agents continuously adapt their visitation policies to evolving importance distributions without requiring a prior map or its evolution, or centralized coordination. Experiments demonstrate robustness across a range of operational adversities: agent attrition, memory constraints, communication delays, and large-scale deployments, outperforming greedy coverage baselines in dynamic settings. Future work will provide theoretical guarantees in time-varying environments and incorporate semantic grounding for heterogeneous robot coordination via foundation models.

 
%

\bibliographystyle{IEEEtran} 
\bibliography{ref}




\end{document}